\documentstyle[preprint,aps,graphicx]{revtex}

\def\red{
}
\def\black{
}

\def\URLtilde{\lower0.2em\hbox{$\tilde{\phantom{a}}$}}
\def\mycomm#1{\hfill\break\strut\kern-3em{\red\tt ====> #1\black}\hfill\break}
\def\mycommNL#1{\strut\kern-3em{\tt ====> #1}\hfill\break}
\def\ds{\displaystyle}

\def\medstrut{\vrule height 2.5ex depth 1.3ex width 0pt}

\catcode`\@=11 % This allows us to modify PLAIN macros.
\def\lsim{\mathrel{\mathpalette\@versim<}}
\def\gsim{\mathrel{\mathpalette\@versim>}}
\def\@versim#1#2{\vcenter{\offinterlineskip
        \ialign{$\m@th#1\hfil##\hfil$\crcr#2\crcr\sim\crcr } }}
\catcode`\@=12 % at signs are no longer letters

\def\sbar{\hbox{$\bar s$}}

\def\thetap{\hbox{$\Theta^+$}}

\def\DPKN{\hbox{$\Delta P_{KN}^2$}}
\def\DPKNave{\hbox{$\medstrut\overline{\Delta P_{KN}^2}$}}
\def\dPKN{\hbox{$\delta P_{KN}^2$}}
\def\eqref#1{(\ref{#1})}

\makeatletter
\def\hlinewd#1{\noalign{\ifnum0=`}\fi
\hrule \@height #1 \futurelet \reserved@a\@xhline}
\def\hwhiteline{\noalign
{\ifnum0=`}\fi\hrule
\@height 0pt\vskip 1.0ex\futurelet \reserved@a\@xhline}
\makeatother
\def\gray{\special{ps: 0.40 setgray}}
\def\black{\special{ps: 0.0 setgray}}

\newcommand{\mydraft}{
\newcount\timecount
\newcount\hours \newcount\minutes  \newcount\temp \newcount\pmhours

\hours = \time
\divide\hours by 60
\temp = \hours
\multiply\temp by 60
\minutes = \time
\advance\minutes by -\temp
\def\hour{\the\hours}
\def\minute{\ifnum\minutes<10 0\the\minutes
    \else\the\minutes\fi}
\def\clock{
\ifnum\hours=0 12:\minute\ AM
\else\ifnum\hours<12 \hour:\minute\ AM
\else\ifnum\hours=12 12:\minute\ PM
    \else\ifnum\hours>12
     \pmhours=\hours
     \advance\pmhours by -12
     \the\pmhours:\minute\ PM
     \fi
    \fi
\fi
\fi
}
\def\fullclock{\hour:\minute}
\begin{centering}
\gray
\font\Hugett  =cmtt12 scaled\magstep4
\hbox{\Hugett Draft:\today,\clock}
\black
\end{centering}
\vskip -1.7cm
$\phantom{a}$
} % end of \draft definition
\def\beq#1{\begin{equation} \label{#1}}
\def\eeq{\end{equation}}

\newskip\humongous \humongous=0pt plus 1000pt minus 1000pt

\newif\ifdtup

\begin{document}
{\tighten
 \preprint {\vbox{
  \hbox{$\phantom{aaa}$}
  \vskip-0.5cm
%\hbox{\today}
%\hbox{}
\hbox{Cavendish-HEP-05/04}
\hbox{TAUP 2794-05}
\hbox{WIS/1/05-JAN-DPP}
\hbox{ANL-HEP-PR-05-5} }}

\title{New tests for experiments producing pentaquarks}
\author{Marek Karliner\,$^{a,b}$\thanks{e-mail: \tt marek@proton.tau.ac.il}
\\
and
\\
Harry J. Lipkin\,$^{b,c}$\thanks{e-mail: \tt
ftlipkin@weizmann.ac.il} }
\address{ \vbox{\vskip 0.truecm}
$^a\;$Cavendish Laboratory, Cambridge University, United Kingdom\\
%\mbox{}\\
and\\
$^b\;$School of Physics and Astronomy \\
Raymond and Beverly Sackler Faculty of Exact Sciences \\
Tel Aviv University, Tel Aviv, Israel\\
\vbox{\vskip 0.0truecm}
$^c\;$Department of Particle Physics \\
Weizmann Institute of Science, Rehovot 76100, Israel \\
and\\
High Energy Physics Division, Argonne National Laboratory \\
Argonne, IL 60439-4815, USA\\
}
\maketitle
%\mydraft
\begin{abstract}%
The distribution of the squared momentum difference $|P_A|^2 - |P_B|^2$ between
the momenta in the laboratory system of two experimentally observed particles
$A$ and $B$ provides a test for whether an observed mass peak  indicates a real
resonance rather than nonresonant background or kinematic reflection. 
The
angular distribution of the relative momenta in the center-of-mass system
exhibits a forward-backward symmetry in the production and decay of any
resonance with a definite parity. This symmetry is not expected in other
nonresonant processes and can be expressed without needing angular
distributions in terms of the easily measured momenta in the laboratory frame
that are already measured and used to calculate the invariant mass of the
system.  Our test 
is especially useful for low statistics experiments where the full
angular distribution cannot be determined. It
can be applied to both fixed-target and collider 
searches for the $\Theta^+$ and $\Theta_c$ pentaquarks.
 \end{abstract}% %} % end tighten

\vfill\eject

The recent experimental discovery
\cite{Nakano:2003qx}
of an exotic 5-quark $K N$ resonance \thetap\
with \hbox{$S={+}1$}, a mass of $\sim$1540 MeV,
a very small width $\lsim 20$
MeV
and a presumed quark configuration $uudd\sbar$  has given rise
to a number of experiments with contrary results \cite{hicks}.

At this point it seems crucial to analyze and
extend both the positive and negative experiments to either establish the
$\Theta^+$ as a real particle and understand this contradiction or to find good
credible reasons against its existence.

Many detailed theoretical pentaquark models have been proposed, but few
address the problem of why certain experiments see it and others do not. We
therefore do not consider them here. The ball is definitely in the experimental
court.  Our purpose is to establish  communication between  theorists who
know which measurements are of theoretical interest and  experimenters who know
which measurements are possible with available facilities.

In this context we note a simple experimental test for the production of any
two-body resonance having a definite parity. The angular distribution of the
relative momentum in the rest frame of the resonance exhibits a
forward-backward symmetry  for the production and decay of a resonance and this
symmetry generally is not present in nonresonant background. We present here a
method to test experimental data for this symmetry while avoiding the
difficulties of measuring angular distributions with poor statistics.

For a simple example of the basic physics, consider a peak arising
in  the invariant mass of the two-particle system of particles
denoted by $A$ and $B$ with masses  $M_A$, and $M_B$ in  a
multiparticle final state. If this is a real resonance with a
definite parity, the angular distribution of the relative momentum
in the $AB$ center-of-mass frame must exhibit a forward-backward
symmetry with respect to any external direction; e.g.  the
direction of the total momentum of the $AB$ system in the
laboratory. We now show how this forward-backward symmetry can be
checked easily by measurements of the magnitudes of the momenta of
the two particles in the laboratory system.

In the $AB$ center-of-mass system which is moving with a velocity
denoted by $\vec v$
with  respect to the laboratory system, the total momentum of the $AB$ system is
zero. Consider the case where the momenta of particles $A$ and $B$ are
perpendicular
to the direction of the incident momentum in the laboratory
system.   In a nonrelativistic approximation the longitudinal components of the
$A$ and $B$ momenta which are zero in the $AB$ center-of-mass system are
just the products of mass and velocity in the laboratory system,

\beq {NR}
{\vec v\over v} \cdot \vec P_A^{NR} = M_A v;
\qquad
{\vec v\over v} \cdot \vec P_B^{NR} = M_B v
\end{equation}
where $\vec P_A^{NR}$ and  $\vec P_B^{NR}$
denote the momenta  respectively of particles $A$ and $B$.
Then the difference
of the momenta squared satisfies
\beq {sumdiffNR}
{{|P_A^{NR}|^2 - |P_B^{NR}|^2}\over{|\vec P_A^{NR} + \vec P_B^{NR}|^2}}=
{{M_A^2 - M_B^2}\over{(M_A + M_B)^2}}
\end{equation}
where we note that the contributions of the transverse components of $A$ and
$B$ momenta which are equal and opposite cancel out in both the numerator and
denominator of the lhs of eq. (\ref {sumdiffNR}).

Eq. (\ref {sumdiffNR}) gives the value of the squared laboratory
momentum difference $|P_A^{NR}|^2 - |P_B^{NR}|^2$ which
corresponds to a resonance decay in which particles $A$ and $B$
are both moving in a direction in the $AB$ center-of mass system
perpendicular to the direction of the total laboratory momentum.
Events having a larger value of  $|P_A^{NR}|^2 - |P_B^{NR}|^2$
correspond to forward emission of particle $A$; a smaller value
corresponds to
 backward
emission. Thus measurements of the laboratory momenta of particles $A$ and $B$
give
information about the angular distribution in the center-of-mass system without
any angular measurements.

A full relativistic treatment of the angular distribution is given below.
However, we note that for the particular case of transverse momenta in the
$AB$ center-of mass system the relativistic corrections to the ratio (\ref
{sumdiffNR}) are simply expressed by replacing the masses $M_A$ and $M_B$ by
the center-of-mass energies $E_A(cm)$ and $E_B(cm)$ and noting that in the
center-of-mass system  the total energy is just the invariant mass $M$ while
the momenta are equal and opposite and cancel out in the squared difference.
Thus the relativistic generalization of eq.  (\ref {sumdiffNR}) is

\beq {sumdiffrel}
{{|P_A|^2 - |P_B|^2}\over{|\vec P_A + \vec P_B|^2}}=
{{E_A(cm)^2 - E_B(cm)^2}\over{[E_A(cm) + E_B(cm)]^2}} =
{{M_A^2 - M_B^2}\over{M^2}}
\end{equation}

We now derive the full relativistic generalization of this  simple
approach and apply it to the particular case of production via a
$K$ or $K^*$ exchange on a nucleon at rest and in the reaction
$K^+ p \rightarrow  \pi^+ \Theta^+ \rightarrow \pi^+ K^+ n
$\cite{LASS}.

In these exchange diagrams for $\Theta^+$ production the reactions at the
baryon vertex are
\beq {barv}
K + N  \rightarrow \Theta^+ \rightarrow K + N ; ~ ~ ~ ~
 K^* + N  \rightarrow \Theta^+ \rightarrow  K + N
\end{equation}

When a $\Theta^+$ spin-1/2 baryon resonance is produced by the
reactions  (\ref {barv}) the angular distribution of the kaon momentum in the
center-of-mass frame of the final $KN$ system is isotropic.
This isotropy produces a  forward-backward symmetry relative to any axis and in
particular the axis defined by the center-of-mass momentum in the laboratory
frame.

This forward-backward symmetry holds for the production and decay of any
resonance with a definite parity.  The amplitudes for two final states
differing only by a reversal of the relative momentum in the center-of-mass
system,  denoted by $\vec p_{cm}$ can differ only by a real phase which depends
upon the parity of the resonance.

 \beq{parity}
A(\vec P,\vec p_{cm}) = \pm A(\vec P,-\vec p_{cm})
\end{equation}
where $\vec P$ denotes the total momentum of the system in the laboratory frame.
On the other hand, if the final $KN$ state is produced by a nonresonant
peripheral reaction like meson exchange, the kaon angular momentum is strongly
peaked forward in the center-of mass system.  The difference
between a symmetric and a forward-peaked distribution can be checked  without
angular measurements by expressing the condition (\ref{parity}) in terms
of the magnitudes of the nucleon and kaon momenta in the laboratory frame.

Let $\vec P_K$, $E_K$, $\vec P_N$ and $E_N$ denote the momenta and energies of
the kaon and nucleon in the laboratory frames. The total momentum and the
momentum difference are four-vectors defined as
\beq {sum_diff}
\begin{array}{ccccc}
\vec P &=&\vec P_K + \vec P_N; \qquad\qquad E &=& E_K + E_N
\\
&&&&\\
%\end{equation}
%\beq {diff}
\vec p &=&\vec P_K - \vec P_N; \qquad\qquad \epsilon &=& E_K - E_N
\end{array}
\end{equation}
The values of the sums and differences in the center-of-mass system are given
by the Lorentz transformation with a velocity $\vec v$
\beq {sumcm}
\vec P_{cm} =\gamma [\vec P - \vec v E] =0; \qquad\qquad
E_{cm} = \gamma [E  - \vec v \cdot \vec P] =M
\end{equation}
\beq {diffcm}
\begin{array}{ccccl}
\vec p_{cm} &=&
\gamma [ \vec p - \vec v \epsilon ]
&=&   \gamma \left[ \ds \vec p -\vec P \cdot {{|P_K|^2 - |P_N|^2 + M_K^2
- M_N^2 }\over{E^2}} \right];
\\
&&&&\\
\epsilon_{cm} &=& \gamma [ \epsilon - \vec v \cdot \vec p]
&=& \gamma \left[ \epsilon - {\ds \vec P \over  \ds \strut E}\cdot \vec p
\,\right]
\end{array}
\end{equation}
where $M$ is the invariant mass of the $KN$ system,
\beq {gamma}
\gamma = {1\over {\sqrt {1 - v^2}}}
\end{equation}
and we have used eq.~(\ref {sumcm}) to obtain and substitute in
eq.~(\ref{diffcm})
\beq {lorentz}
\vec v \epsilon = {\vec P \over  E} \epsilon = {\vec P \over  E^2}
\cdot  \epsilon E  = \vec P
\cdot {{E_K^2 - E_N^2}\over{E^2}} = \vec P
\cdot {{|P_K|^2 - |P_N|^2 + M_K^2
- M_N^2 }\over{E^2}}
\end{equation}

To express the condition (\ref{parity}) in terms of the laboratory momenta
$\vec P_K$, $\vec P_N$ and kaon, nucleon and resonance masses denoted
respectively by $M_K$, $M_N$ and $M$, we note that
\beq {90deg}
\begin{array}{ccc}
\vec p_{cm} \cdot \vec P &=&
\gamma \left[ \vec p \cdot \vec P
- |P|^2\cdot \ds { |P_K|^2 - |P_N|^2 + M_K^2 - M_N^2 \over E^2}
\right]  =
\\
&&\\
&=&
\gamma \left [M^2 \cdot \ds {|P_K|^2 - |P_N|^2 \over E^2} -  |P|^2\cdot{{M_K^2
- M_N^2}\over{E^2}}\right]
\end{array}
\end{equation}
where we used $\vec p\cdot \vec P = |P_K|^2 - |P_N|^2$. The
squared momentum difference \hbox{$|P_K|^2 - |P_N|^2\equiv\DPKN$}
is thus given by
\beq{90deg1}
\DPKN =
 {{ M_K^2 - M_N^2}\over{M^2}}\cdot   |P|^2
 + {E^2\over \gamma M^2} \,\, \vec p_{cm}
\cdot \vec P \equiv \DPKNave + {E^2\over \gamma M^2} \,\, \vec
p_{cm} \cdot \vec P
\end{equation}
where $\DPKNave=( M_K^2 - M_N^2)\cdot |P|^2/M^2 $\,.

The condition (\ref{parity}) implies that for any given total momentum value
\hbox{$\vec P = \vec P_1$}
the counting rate observed at a value of  the squared
momentum difference
\hbox{$\DPKN=\DPKNave + \dPKN$}
will be equal to the counting rate observed at
\hbox{$\DPKN=\DPKNave - \dPKN$}
\beq {parity2}
N(\DPKN= \DPKNave + \dPKN) =
N(\DPKN= \DPKNave - \dPKN)
\end{equation}
The mean value of the squared momentum difference $|P_K|^2 - |P_N|^2$ is given
by
\beq {mean}
\langle|P_K|^2 - |P_N|^2\rangle  = {{ M_K^2 - M_N^2 }\over{M^2}}\cdot   |P|^2
= \DPKNave
\end{equation}
and the distribution of the squared momentum difference $|P_K|^2 - |P_N|^2$ is
symmetric around the mean value (\ref{mean}).

This mean is negative and proportional to the kaon-nucleon mass difference,
because  a boost with the same velocity for the kaon and nucleon increases the
nucleon momentum more than the kaon momentum. A higher value of $P_K^2 - P_N^2$
corresponds to  forward scattering, a lower to backward scattering. Thus a
forward-peaked background will show up with higher values of the difference
between the kaon and nucleon momenta $P_K^2 - P_N^2$ in the  laboratory
system.

For the case of an isotropic angular distribution in the center of mass system
and a very narrow resonance, $\vec p_{cm}$ is constant in magnitude and

\beq{isotrop}
\vec p_{cm} \cdot \vec P = |\vec p_{cm}| \cdot |\vec P| \cos \theta
\end{equation}
where $\theta$ is the angle between $\vec p_{cm}$ and $\vec P$. The
distribution of the squared momentum difference $|P_K|^2 - |P_N|^2$
is flat between the limits corresponding to $\cos \theta = \pm 1$.

\beq {isotrop2}
\DPKNave
-{E^2\over\gamma M^2}\,\, |\vec p_{cm}| \cdot |\vec P|
\,\,\leq\,\,
\DPKN
\,\,\leq\,\,
\DPKNave
+{E^2\over\gamma M^2}\,\, |\vec p_{cm}| \cdot |\vec P|
\end{equation}

The question now arises whether eqs.~(\ref {parity2}) and (\ref{mean}) can
provide useful information in real data with  all the acceptance restrictions
and give a simple test to see whether it works at all.

We first note that alternative mechanisms that have been suggested using
kinematic reflections to explain the $\Theta^+$ mass peak\cite{dzierba} will
generally not have the forward-backward asymmetry nor satisfy
eqs.~(\ref{parity2}) and (\ref{mean}). Comparing their predicted squared
momentum difference $|P_K|^2 - |P_N|^2$ distributions with measured data can
provide additional checks on these alternatives.

Further investigation of possible uses with real data raise two questions:

\begin{enumerate}
\item
Is there a significant difference between the angular distributions of
signal and background events?
\item
If the answer to (1) is yes, is the difference still significant when
only events that meet the detector acceptance are included?
\end{enumerate}

If the answers to (1) and (2) are yes, there may be ways to improve the
signal/background ratio by cutting out events which are mainly background.

     For a simple test to see whether this makes sense at all in a real
experiment with real detector acceptance limitations, consider the
following.

     Separate the events into two bins, with half of the events in each.
Put the events with the highest values of kaon momentum in one bin, those
with the lowest values in the other,

     Now plot the mass distributions separately for each of the two bins. The
signals from a resonance are expected to be equal in the two by
eq.~(\ref{parity2}).  But if the background is mainly peaked forward in the
center of mass system, there should be more background events in the bin with
the higher kaon momentum.

     If the two mass distributions turn out to be the same, there is no
point in following this further. But if the two mass distributions are
different, it will be worth while trying to develop this approach further.

The basic idea suggesting that there might be a difference is the
assumption that the kaons going forward in the center-of-mass
frame are mainly background. These will be the kaons having the highest
momentum in the laboratory system. If this is true, the signal to
background ratio can be improved by cutting out the events with the
highest kaon momentum.

If there is forward peaking, indicating a strong nonresonant background,
cuts
removing the events having the most positive values of \DPKN\ can
eliminate the background coming from strongly forward events,
without excessively harming the resonant signal.

In many experiments where the $\Theta^+$ resonance is produced, other
well-known resonances are also produced. In particular the $\Lambda(1520)$
resonance has often been used for comparison with the $\Theta^+$\cite{hicks}.
The momentum  distributions of such final states of  other resonances like the
$\Lambda(1520)$ must certainly satisfy the conditions  (\ref {parity2}) and
(\ref{mean}). Any deviations from these conditions must be due to variations in
the detector acceptance as a function of the momenta. These can provide useful
information on the detector acceptances for the  $\Theta^+$ data. In the
particular case of the $\Lambda(1520)$, the CLAS data\cite{hicks} show a very
strong peak with comparatively low background. Measuring the  squared
momentum difference $|P_K|^2 - |P_N|^2$ distributions both in the resonance
peaks and in the backgrounds on both sides of the resonance can provide
interesting insight into the applicability of this method.

The conditions (\ref {parity2}) and (\ref{mean}) hold for any experiment in
which a resonance is produced, not only in production from a target at
rest. 
They may not be useful if the background also satisfies these
conditions and does not have a strong forward or backward peaking. However they
may still supply useful checks and information in all cases.

We believe our approach
is especially useful for low statistics experiments where the full
angular distribution cannot be determined. It
can be applied to both fixed-target experiments, such as photoproduction,
$pp$, $pA$, $KN$, $KA$ and $\nu A$ collisions,
as well as $eN$ and $e^+e^-$ collider searches 
for the $\Theta^+$ \cite{hicks} and $\Theta_c$
\cite{Aktas:2004qf}-\kern-0.3em\cite{ThetacSearches}
pentaquarks.

\section*{Acknowledgements}

The research of one of us (M.K.) was supported in part by a grant from the
Israel Science Foundation administered by the Israel
Academy of Sciences and Humanities.
The research of one of us (H.J.L.) was supported in part by the U.S. Department
of Energy, Division of High Energy Physics, Contract W-31-109-ENG-38.
We thank
Uri Karshon,
Takashi Nakano,
Jim Napolitano,
Jin Shan,
Igor Strakovsky,
and
San Fu Tuan
for discussions and comments.

%----------------------------------------------------------------------
% This prevents REFERENCES from forcing a page break
%\def\newpage{\vskip10ex}
%
\catcode`\@=11 % This allows us to modify PLAIN macros
\def\references{
\ifpreprintsty \vskip 10ex
%\ifpreprintsty \newpage
%
\hbox to\hsize{\hss \large \refname \hss }\else
\vskip 24pt \hrule width\hsize \relax \vskip 1.6cm \fi \list
{\@biblabel {\arabic {enumiv}}}
{\labelwidth \WidestRefLabelThusFar \labelsep 4pt \leftmargin \labelwidth
\advance \leftmargin \labelsep \ifdim \baselinestretch pt>1 pt
\parsep 4pt\relax \else \parsep 0pt\relax \fi \itemsep \parsep \usecounter
{enumiv}\let \p@enumiv \@empty \def \theenumiv {\arabic {enumiv}}}
\let \newblock \relax \sloppy
 \clubpenalty 4000\widowpenalty 4000 \sfcode `\.=1000\relax \ifpreprintsty
\else \small \fi}
\catcode`\@=12 % at signs are no longer letters
%-----------------------------------------------------------------
%{\tighten

} % end of global \tighten
\end{document}